\documentclass[twocolumn]{article}
\usepackage{amsmath}
\usepackage{algorithm}
\usepackage{algpseudocode}
\usepackage{amsthm}
\usepackage{graphicx}
\usepackage{appendix}
\usepackage{balance}

\usepackage[utf8]{inputenc} % allow utf-8 input
\usepackage[T1]{fontenc}    % use 8-bit T1 fonts
\usepackage{hyperref}       % hyperlinks
\hypersetup{
    colorlinks=true,
    linkcolor=blue,
    citecolor=blue,
    filecolor=magenta,
    urlcolor=blue,
}
\usepackage{url}            % simple URL typesetting
\usepackage{booktabs}       % professional-quality s
\usepackage{amsfonts}       % blackboard math symbols
\usepackage{nicefrac}       % compact symbols for 1/2, etc.
\usepackage{microtype}      % microtypography

\usepackage{tabularx}
\usepackage{booktabs}
\usepackage[]{natbib}
\usepackage{enumitem}
\usepackage{graphicx}
\usepackage{array}
\usepackage{ulem}
\usepackage{amssymb}
\usepackage{amsmath}
\usepackage{fancyhdr}
\usepackage{float}  %float allows [H] to force a Exhibit or  where you want it
\usepackage{listings}
\usepackage{soul}

\usepackage{newfloat}
\DeclareFloatingEnvironment[%
   fileext=loe,
   name=Exhibit
]{exhibit}

\AtBeginEnvironment{thebibliography}{\sloppy\hbadness=10000}
\begin{document}

\title{Multivariate Financial Forecasting using the \\Chronos Time Series Foundation Models\thanks{Das is a professor of Finance and Data Science, and Goyal and Yadav are graduate students in the Leavey School of Business at Santa Clara University.}}
\author{Sanjiv R. Das, Tarang Goyal, Mohini Yadav \\ Santa Clara University\\ \texttt{srdas@scu.edu; tgoyal@scu.edu; myadav@scu.edu}}

% \title{Multivariate Financial Forecasting using the \\Chronos Time Series Foundation Models}

\maketitle

\begin{abstract}
Using Chronos-2, an open-source time-series foundation model, we evaluate pretrained time-series models for economic and financial forecasting with an emphasis on whether multivariate (MV) inputs improve accuracy relative to univariate (UV) baselines. The study covers two panels---the Magnificent-7 equities and U.S. Treasury interest rates---as well as a combined panel, using rolling monthly evaluations from 2000--2025. We vary input window lengths and forecast horizons and report RMSE and MAPE. Across datasets, MV forecasts consistently outperform UV forecasts, with especially strong gains for interest rates and meaningful improvements for equities. Series-level comparisons show MV improvements in every case, and error dispersion is generally lower under MV inputs. We also provide parameter-heatmap and time-series visualizations. However, mixing time series across equity and interest rate markets reduces forecast accuracy, indicating that adding noisy context degrades model performance. Overall, the results indicate that foundation models can leverage cross-series information to improve forecast accuracy in finance, and that the benefits are strongest when related series are modeled jointly under disciplined rolling protocols. Other than using an open-source foundation model, this paper also  showcases how AI may be used for financial research. 
\end{abstract}

\section{Introduction}

Time-series forecasting in finance remains a contested area for modern machine learning. While deep learning has transformed natural language processing and computer vision, its advantage in economic and financial forecasting is far less clear. Recent comparative studies of U.S. Treasury yield curve forecasting report that classical econometric baselines and simple benchmarks can outperform many modern machine-learning models across long spans of daily data, with only a few learning approaches delivering competitive gains. These findings underscore an open question for practitioners: when, and under what conditions, do multivariate forecasting methods provide reliable improvements over univariate forecasting? Adding data for a forecast brings up the standard question of whether it introduces signal or noise.

The emergence of foundation models---large, self-supervised models trained on broad data and adapted to downstream tasks---has begun to reshape time-series forecasting. The promise of these models is not only scale, but also transfer: a single model can exploit shared structure across diverse series and potentially improve predictive accuracy in domains where data are limited or noisy. Yet evidence from forecasting competitions and applied finance \citep{singh_yield_2025} suggests that pure deep-learning solutions often struggle unless they are hybridized or carefully engineered.

This paper evaluates whether foundation-model forecasting can deliver practical improvements in economic and financial settings, focusing on Chronos-2, an open-source time-series foundation model (GitHub: \url{https://github.com/amazon-science/chronos-forecasting}). Just as large language models generate a sequence of words, Chronos-2 Chronos-2 is trained on thousands of time series, and generates forward sequences of data based on multivariate histories. Our central hypothesis is that multivariate (MV) inputs, when properly chosen, aligned, and modeled, can help the model extract cross-series information and outperform univariate (UV) baselines. We test this hypothesis across two economically meaningful panels: the Magnificent-7 equities and a set of U.S. Treasury interest rates, and we also study a combined panel to assess whether joint modeling improves results.

The contribution of the paper is fourfold. First, we provide a systematic comparison of MV versus UV forecasting across equities and rates with a consistent rolling evaluation protocol. We find that the MV approach beats the UV one. Second, we examine whether MV gains differ by asset class and whether joint stock and interest rates modeling improves accuracy. In this case, we find that adding a different class of time series does not improve forecasting, highlighting that expanding the basis set for modeling does not always improve forecast accuracy. Indeed, there is a signal to noise tradeoff that applies here. Third, we offer large-scale outputs---error metrics, forecasts with uncertainty quantiles, and visualization-ready artifacts---to support robust evaluation and downstream use. Fourth, we provide in Appendix \ref{ai_workflow} the process by which AI was not only the subject of the paper, but was also used to produce the paper. Collectively, these results clarify the conditions under which foundation models add value relative to standard forecasting practice in finance.

\subsection{Brief Literature Review}

The literature on financial time-series forecasting spans classical econometrics, machine learning, and recent deep-learning approaches. Standard time-series tools and statistical benchmarks remain central in practice and are documented in foundational texts \citep{shumway_time_2025}. In fixed-income applications, yield curve forecasting has been studied using no-arbitrage and factor-augmented approaches \citep{moench_forecasting_2006} as well as forecast combinations \citep{caldeira_predicting_2013}. Machine-learning methods have been applied to yield curves with mixed results, including statistical ML approaches \citep{sambasivan_statistical_2017} and more recent comparative studies that find strong performance for classical econometric baselines relative to many ML and deep-learning models \citep{singh_yield_2025}. Related work on bond risk premia also highlights the potential and limitations of ML for term-structure prediction \citep{bianchi_bond_2020,hoogteijling_forecasting_2021}.

For equities and broader financial markets, surveys emphasize the breadth of deep-learning adoption but also the variability of performance across datasets and tasks \citep{sezer_financial_2019}. Empirical studies comparing ARIMA, LSTM, and Prophet illustrate that model choice can be highly data-dependent, with traditional statistical methods remaining competitive in many settings \citep{sunki_time_2024}. The broader financial machine-learning literature argues for careful feature engineering, rigorous validation, and robust benchmarks to avoid overfitting and spurious gains \citep{lopez_de_prado_advances_2018}. These themes motivate careful UV baselines and out-of-sample evaluation in this paper.

Recent advances in deep-learning forecasting include specialized architectures such as NHITS \citep{challu_nhits_2023} and, more recently, foundation models for time series. A recent review is available in \cite{zhang_review_2026}. Chronos is a large, self-supervised forecasting model trained on diverse datasets and designed for transfer across domains \citep{ansari_chronos_2024}. This paper contributes to the growing evidence on foundation models by testing whether MV inputs can systematically improve forecasting accuracy in economically important panels of equities and rates, and by reporting results using consistent rolling protocols and multiple error metrics. In this paper we focus on the newest generation in this class of models, namely Chronos-2 \citep{ansari_chronos-2_2025}.

\section{Methodology}

This section develops the research questions and the experimental framework including the Chronos-2 forecasting pipeline. We describe the data scope, model inputs and horizons, rolling evaluation design, and the artifacts produced for analysis and downstream use.

\subsection{Research Questions}
The study addresses three related questions. First, we test whether multivariate (MV) forecasting yields more accurate predictions than univariate (UV) methods. By restricting attention within a single model, Chronos-2, we are able to focus specifically on this effect in the class of foundation models. Second, we examine whether MV gains differ across asset classes by comparing stocks to interest rates. Third, we evaluate whether forecasting stocks and rates jointly improves accuracy relative to modeling each group separately. 

\subsection{Overview of the Chronos-2 model}

Chronos-2 \citep{ansari_chronos-2_2025} extends the foundation-model approach to time-series forecasting by moving beyond purely univariate inputs toward a universal, zero-shot framework that can handle multivariate groups and covariate-informed tasks. The motivation is pragmatic: real forecasting problems often involve co-evolving series or external drivers, yet most pretrained models are limited to single-series histories. Chronos-2 addresses this gap by enabling joint inference across related series without task-specific retraining, simplifying deployment and improving transfer to new domains.

Architecturally, Chronos-2 introduces a group-attention mechanism that supports in-context learning across sets of series. A ``group'' can represent variates of a multivariate series, multiple related targets, or targets plus covariates. The attention design enables efficient information sharing and allows the model to learn cross-series structure at inference time, which is crucial for multivariate forecasting and covariate conditioning.

To train these capabilities, Chronos-2 is built on synthetic multivariate constructions derived from univariate time series, enforcing diverse dependence structures during pretraining. Evaluation is reported on multivariate and covariate-aware benchmarks such as fev-bench, GIFT-Eval, and Chronos Benchmark II, where Chronos-2 attains state-of-the-art performance, particularly on covariate-conditioned tasks. These design choices align with our use case: leveraging joint structure across equities and rates and assessing whether MV inputs yield systematic gains over UV baselines.

The methodology behind Chronos-2 centers on transforming time series forecasting into a universal, zero-shot task through a specialized transformer architecture and innovative training strategies. Unlike its predecessor model, which was primarily univariate, Chronos-2 leverages in-context learning (ICL) to handle multivariate data and covariates without needing task-specific fine-tuning. Many specific techniques are deployed in pre-training this model (see Figure 1 in the Chronos-2 technical report). 

\begin{enumerate}

\item Tokenization and Robust Scaling. 
The pipeline begins by normalizing input data using a robust scaling scheme. This involves standardizing the series and applying a $sinh^{-1}$ transformation to stabilize variance and reduce the impact of extreme outliers. Categorical covariates are converted to real-valued representations using target or ordinal encoding. After scaling, the data is augmented with meta features, including a relative time index and a binary mask that indicates observed versus missing values.

\item Patch-Based Representation. 
To process the data efficiently, Chronos-2 uses patching, where the continuous time series and its meta features are divided into non-overlapping segments of length $P$. These patches are mapped into high-dimensional embeddings via a residual network. A unique REG token is inserted between historical context and future patches to serve as an attention sink and separator.

\item Core Architecture: Time and Group Attention
The model is an encoder-only transformer that alternates between two critical attention layers: (a) Time Attention: This layer aggregates information across temporal patches within a single time series, utilizing Rotary Position Embeddings (RoPE) to maintain temporal order. (b) Group Attention: This is the core innovation of Chronos-2. It allows the model to share information across different time series within a ``group'' at each patch index. A group can flexibly represent variates of a multivariate series, a target with its covariates, or even a collection of related univariate series for cross-learning.

\item Multivariate and Covariate Integration. 
Chronos-2 manages different task types---univariate, multivariate, and covariate-informed---within a unified framework by assigning Group IDs. For multivariate tasks, all related variates share an ID, enabling the group attention layer to model their dependencies. When covariates are present, they are included in the group, and known future values are provided in the future input matrix $W$, allowing the model to condition its forecasts on them.

\item Training and Quantile Forecasting.
The model is trained in two stages to handle varying context lengths, starting at 2048 steps and extending to 8192. It uses a quantile regression objective to produce probabilistic forecasts across 21 quantiles (from 0.01 to 0.99), providing a comprehensive view of uncertainty and rare event risks. Remarkably, its advanced ICL capabilities are largely learned from synthetic datasets that impose complex multivariate structures on univariate base series.

\end{enumerate}

In sum, the methodology of Chronos-2 involves an encoder-only transformer architecture that utilizes a novel group attention mechanism and robust scaling to perform zero-shot univariate, multivariate, and covariate-informed forecasting by sharing information across related time series in-context.

\subsection{Implementation}

The data consist of three panels: the Magnificent-7 stocks (AAPL, AMZN, GOOGL, MSFT, NFLX, NVDA, TSLA) ($K=7$), Treasury interest rates for ten different maturities ($K=10$), and a combined panel of all series ($K=17$). This design supports both within-group and cross-group forecasting comparisons.

We use input window lengths $n \in \{126, 252, 504, 756\}$ rading days, and forecast horizons $m \in \{21, 63\}$ trading days to cover short to medium histories (0.5 to 3 years) and near- to mid-term prediction horizons (1 to 3 months).

The evaluation period spans 2000--2025 with monthly rolling forecasts to reflect realistic re-estimation and deployment cadence. Forecast accuracy is measured using RMSE (root mean squared error) and MAPE (mean absolute percentage error) to capture both scale-dependent and scale-free error behavior. 

To assess temporal robustness, we compare performance across pre-2023 and post-2023 regimes using a training cutoff that separates the sample. We will also argue that this is a check on information leakage in the model. 

\section{Results}

\subsection{Forecasts within asset classes}

One of the goals of the paper is to examine the extent of improvement when multivariate forecasting (MV) is used instead of univariate forecasting (UV). Table~\ref{tab:avg-performance} summarizes average performance by dataset and mode. Results are aggregated across all tickers, and in-sample and out-sample combinations. MV forecasts outperform UV for both rates and stocks, with large MAPE and RMSE reductions in rates and lesser gains in stocks. The standard deviation of the forecast error is also smaller for MV versus UV forecasts. Overall, the results suggest that multivariate models are easily implemented in Chronos-2 and provide robust forecasts. 

\begin{table}[H]
\centering
\caption{\label{tab:avg-performance} \small Average performance by dataset and mode. Lower mean error across all experiments (for values of input length $n=\{126, 252, 504, 756\}$, and forecast period $m=\{21, 63\}$) measured by MAPE and RMSE is better. Data for the rolling periods is from 2000 through 2025. Predictions periods begin 3 years after the start of the dataset. }
\resizebox{\columnwidth}{!}{%
\begin{tabular}{llrrrr}
\toprule
Dataset & Mode & \multicolumn{2}{c}{MAPE} & \multicolumn{2}{c}{RMSE} \\
 &  & mean & std & mean & std \\
\midrule
Rates  & MV & 0.0497 & 0.1673 & 0.0418 & 0.0445 \\
Rates  & UV & 0.1233 & 0.2213 & 0.1865 & 0.1783 \\
Stocks & MV & 0.0706 & 0.0728 & 3.9619 & 9.0440 \\
Stocks & UV & 0.0844 & 0.0834 & 5.0395 & 11.5710 \\
\bottomrule
\end{tabular}
}
\end{table}

Table~\ref{tab:uv-mv-comparison} reports UV vs.\ MV results by series. MV improves both MAPE and RMSE for every rate and stock. This is a clear win for the MV forecasting model, suggesting that the information from additional related time series adds value when used in the realm of a pre-trained foundation time series model. 

\begin{table*}[t]
\centering
\caption{\label{tab:uv-mv-comparison} \small UV vs.\ MV comparison by series (lower is better). Improvements are mean UV error minus mean MV error. Results are averaged across values of input length $n=\{126, 252, 504, 756\}$, and forecast period $m=\{21, 63\}$. Data for the rolling periods is from 2000 through 2025. Predictions periods begin 3 years after the start of the dataset. 
}
\resizebox{\textwidth}{!}{%
\begin{tabular}{llrrrrrr}
\toprule
Dataset & Series & MAPE\,(MV) & MAPE\,(UV) & RMSE\,(MV) & RMSE\,(UV) & MAPE\,Imp. & RMSE\,Imp. \\
\midrule
Rates & DGS3MO & 0.2355 & 0.3186 & 0.0618 & 0.1227 & 0.0831 & 0.0609 \\
Rates & DGS6MO & 0.0833 & 0.1770 & 0.0455 & 0.1257 & 0.0937 & 0.0802 \\
Rates & DGS1   & 0.0572 & 0.1326 & 0.0454 & 0.1424 & 0.0754 & 0.0970 \\
Rates & DGS2   & 0.0334 & 0.1270 & 0.0413 & 0.1845 & 0.0936 & 0.1432 \\
Rates & DGS3   & 0.0243 & 0.1181 & 0.0388 & 0.2068 & 0.0938 & 0.1680 \\
Rates & DGS5   & 0.0171 & 0.0972 & 0.0359 & 0.2232 & 0.0801 & 0.1873 \\
Rates & DGS7   & 0.0134 & 0.0827 & 0.0345 & 0.2278 & 0.0693 & 0.1933 \\
Rates & DGS10  & 0.0114 & 0.0703 & 0.0344 & 0.2184 & 0.0589 & 0.1840 \\
Rates & DGS20  & 0.0099 & 0.0572 & 0.0364 & 0.2120 & 0.0473 & 0.1756 \\
Rates & DGS30  & 0.0113 & 0.0520 & 0.0442 & 0.2012 & 0.0407 & 0.1570 \\
\midrule
Stocks & AAPL  & 0.0599 & 0.0720 & 3.3016  & 4.1563  & 0.0121 & 0.8547 \\
Stocks & AMZN  & 0.0567 & 0.0707 & 3.0825  & 4.0517  & 0.0140 & 0.9692 \\
Stocks & GOOGL & 0.0478 & 0.0635 & 3.1803  & 4.0070  & 0.0157 & 0.8267 \\
Stocks & MSFT  & 0.0368 & 0.0466 & 4.2003  & 6.1519  & 0.0098 & 1.9516 \\
Stocks & NFLX  & 0.1040 & 0.1145 & 1.7871  & 2.0192  & 0.0105 & 0.2321 \\
Stocks & NVDA  & 0.0906 & 0.1107 & 1.5994  & 1.8420  & 0.0201 & 0.2426 \\
Stocks & TSLA  & 0.1098 & 0.1242 & 13.6796 & 16.7870 & 0.0144 & 3.1074 \\
\bottomrule
\end{tabular}
}
\end{table*}

\begin{figure*}[t]
\centering
{\bf Rates forecast MAPE}\\
\includegraphics[width=\textwidth]{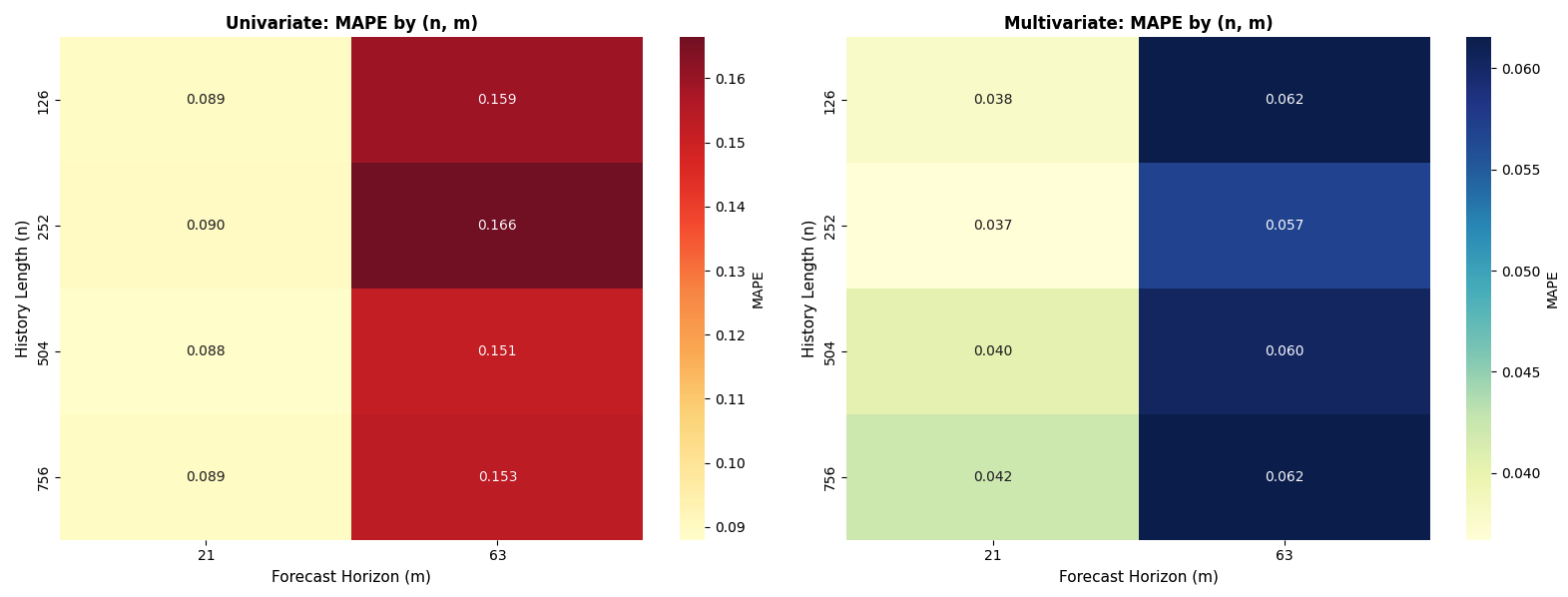}
{\bf Stocks forecast MAPE}\\
\includegraphics[width=\textwidth]{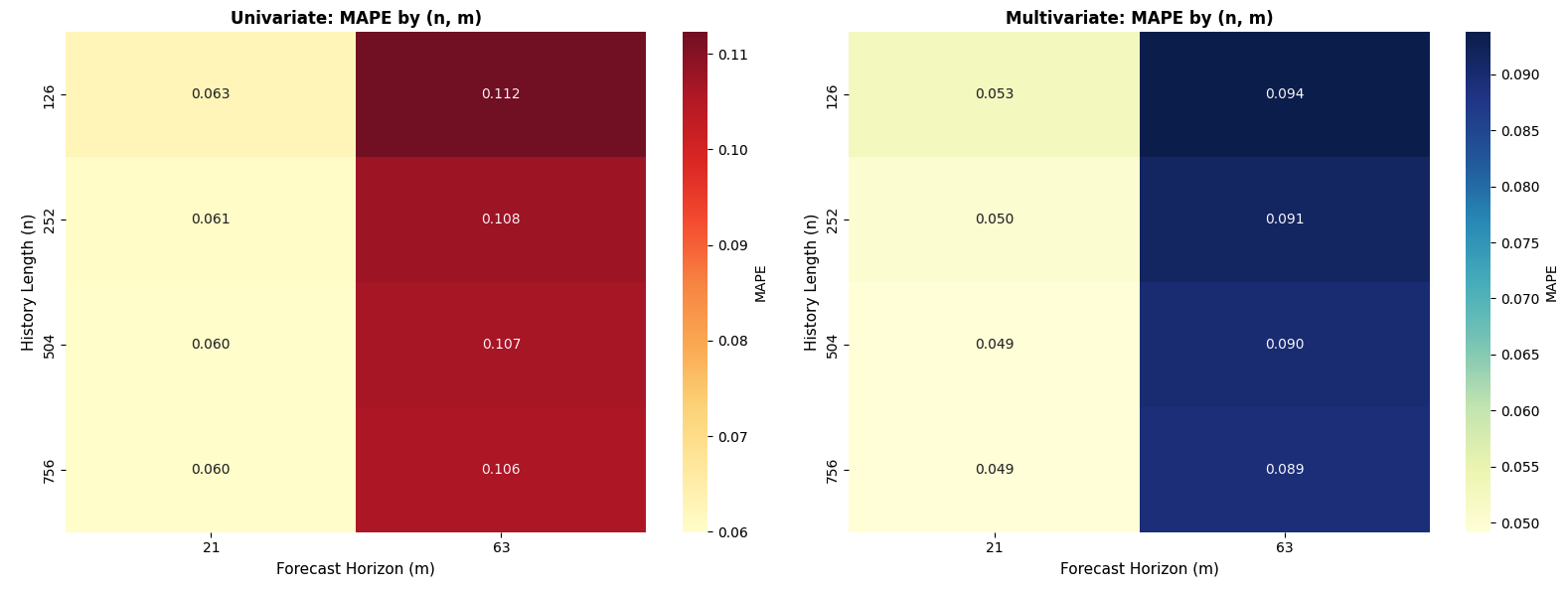}
\caption{\label{fig:parameter_heatmap} \small Parameter heatmap comparing MV and UV forecasting accuracy across window lengths and horizons (based on MAPE). The upper plot is for Rates and the lower one is for Stocks. In each plot, the left side is for the 21 working day forecast (1 month forecast) and the right side is for the 63 working day forecast (3 month forecast)}
\end{figure*}

Figure~\ref{fig:parameter_heatmap} shows the breakdown of errors by the length $n$ of the input time series, and the length $m$ of the forecast series. For both rates and stocks, the forecast error is smaller for the MV approach than for the UV approach. As expected the forecast error is greater for the 63 day forecast than for the 21 day one. But the input series length $n$ does not seem to matter, accuracy levels are not that different across input length, which varies from half a year to three years.

\begin{figure*}[t]
\centering
\includegraphics[width=\textwidth]{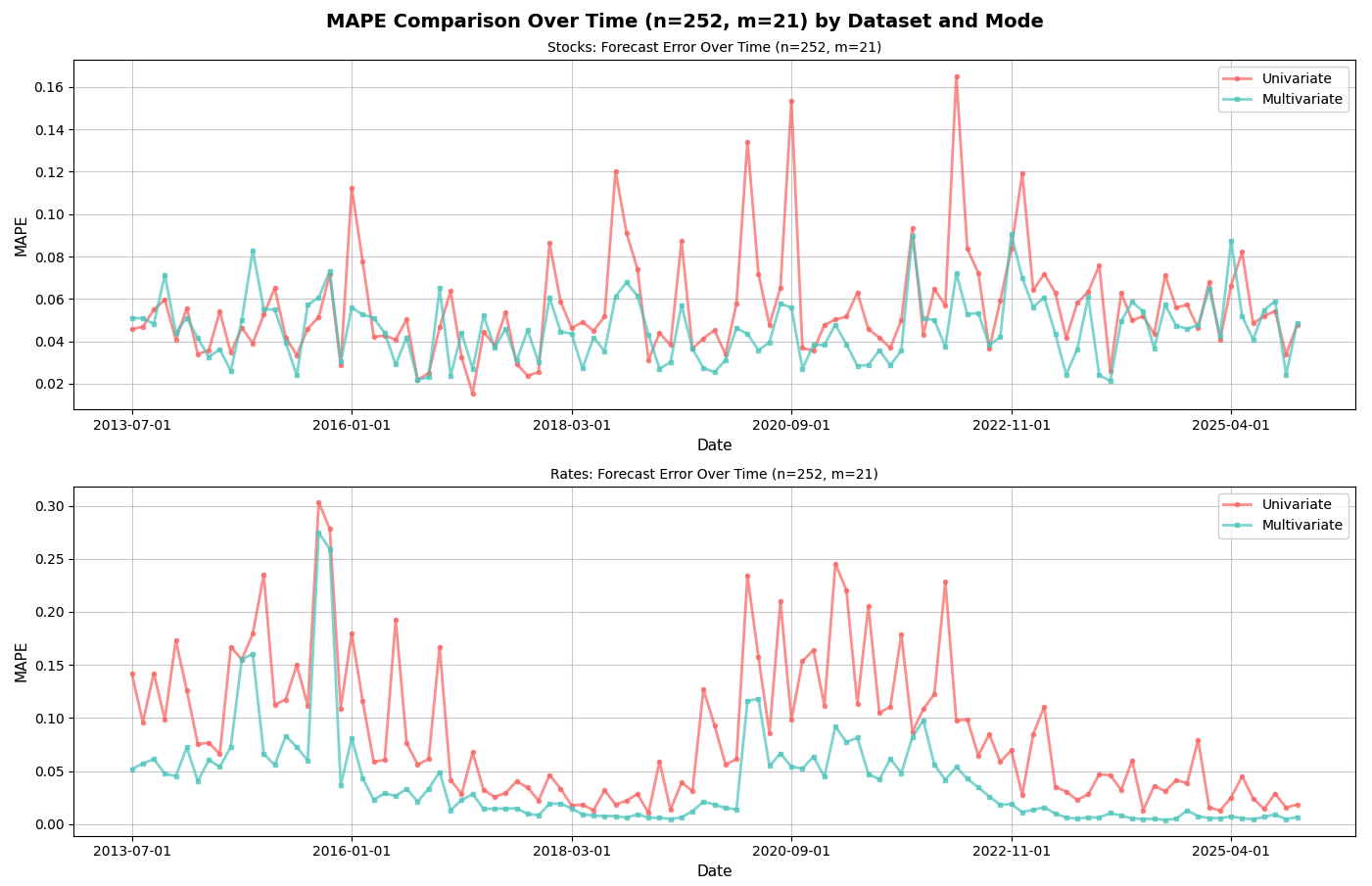}
\caption{\label{fig:uv_mv_time_series} \small Time series of MV and UV MAPE errors over the evaluation period. This is for the dataset running from July 2010 through December 2025. }
\end{figure*}

The time series of MAPE UV and MV errors is shown in Figure~\ref{fig:uv_mv_time_series}. Across all experimental conditions, It is clear that the errors are greater for the UV series for both rates and stocks, with this being more pronounced for interest rates.  In both series the errors seem to be higher when the rates or stocks are at higher levels, even though these are percentage errors. Overall, Figures~\ref{fig:parameter_heatmap} and \ref{fig:uv_mv_time_series} show that the MV forecasts are more accurate than the UV ones.

\subsection{Forecasts with combined asset classes}

In this section we consider forecasts using data from both the Rates and Stocks datasets combined. This so-called small ``world'' model is assessed to see how cross-domain information may be used to improved forecasts. To be more specific, each rate and stock times series is now forecast using the combined information of rates and stock data. This assesses whether there is additional information that may be exploited across the two families of time series. 

The combination of both datasets is only undertaken for the period July 2010 through December 2025, as all stock series are not consistently available through this entire period. Results are shown in Table \ref{tab:uv-mv-combined}. A comparison of the top half of the table with the bottom half shows that the additional data series from another market does not improve forecasts. If anything, this seems to slightly degrade the forecast quality, possibly adding noise instead of additional signal. In the realm of foundation models we can think of this as adding context that distracts the model from the domain in which the forecast is being made.

\begin{table*}
\centering
\caption{\label{tab:uv-mv-combined} \small Combined dataset forecasts. UV vs.\ MV comparison by series (lower is better). Improvements are mean UV error minus mean MV error. Results are averaged across values of input length $n=\{126, 252, 504, 756\}$, and forecast period $m=\{21, 63\}$. Data is for the period July 2010 through December 2025, the period when there is complete data for all stocks and rates. For comparison, we show the results when forecasts are based on the individual datasets for Rates and Stocks, as well as when forecasts are based on the combination of both datasets. Predictions periods begin 3 years after the start of the dataset. (As expected the UV forecast errors do not change.) 
}
\resizebox{\textwidth}{!}{%
\begin{tabular}{llrrrrrr}
\toprule
Dataset & Series & MAPE\,(MV) & MAPE\,(UV) & RMSE\,(MV) & RMSE\,(UV) & MAPE\,Imp. & RMSE\,Imp. \\
\midrule
Rates & DGS3MO & 0.2304 & 0.2849 & 0.0597 & 0.1052 & 0.0545 & 0.0455 \\
Rates & DGS6MO & 0.0937 & 0.1714 & 0.0500 & 0.1179 & 0.0777 & 0.0679 \\
Rates & DGS1 & 0.0609 & 0.1289 & 0.0475 & 0.1363 & 0.0680 & 0.0888 \\
Rates & DGS2 & 0.0366 & 0.1216 & 0.0426 & 0.1727 & 0.0850 & 0.1301 \\
Rates & DGS3 & 0.0257 & 0.1111 & 0.0378 & 0.1883 & 0.0854 & 0.1505 \\
Rates & DGS5 & 0.0185 & 0.0959 & 0.0341 & 0.2025 & 0.0774 & 0.1684 \\
Rates & DGS7 & 0.0144 & 0.0843 & 0.0321 & 0.2086 & 0.0699 & 0.1765 \\
Rates & DGS10 & 0.0128 & 0.0759 & 0.0325 & 0.2045 & 0.0631 & 0.1720 \\
Rates & DGS20 & 0.0116 & 0.0628 & 0.0353 & 0.1993 & 0.0512 & 0.1640 \\
Rates & DGS30 & 0.0122 & 0.0573 & 0.0396 & 0.1923 & 0.0451 & 0.1527 \\
\midrule
Stocks & AAPL  & 0.0500 & 0.0621 & 5.6015  & 7.1124  & 0.0121 & 1.5109 \\
Stocks & AMZN  & 0.0462 & 0.0617 & 5.2798  & 7.0294  & 0.0155 & 1.7496 \\
Stocks & GOOGL & 0.0411 & 0.0545 & 4.8211  & 6.0685  & 0.0134 & 1.2474 \\
Stocks & MSFT  & 0.0305 & 0.0453 & 6.6967  & 10.1334 & 0.0148 & 3.4367 \\
Stocks & NFLX  & 0.0792 & 0.0902 & 3.0946  & 3.4715  & 0.0110 & 0.3769 \\
Stocks & NVDA  & 0.0743 & 0.0980 & 2.7598  & 3.3253  & 0.0237 & 0.5655 \\
Stocks & TSLA  & 0.1090 & 0.1258 & 16.6789 & 20.7258 & 0.0168 & 4.0469 \\
\bottomrule
Dataset & Series & MAPE\,(MV) & MAPE\,(UV) & RMSE\,(MV) & RMSE\,(UV) & MAPE\,Imp. & RMSE\,Imp. \\
\midrule
Combined & DGS3MO & 0.2392 & 0.2849  & 0.0655  & 0.1052  & 0.0457  & 0.0397  \\
Combined & DGS6MO & 0.1002 & 0.1714  & 0.0543  & 0.1179  & 0.0712  & 0.0636  \\
Combined & DGS1   & 0.0671 & 0.1289  & 0.0518  & 0.1363  & 0.0618  & 0.0845  \\
Combined & DGS2   & 0.0392 & 0.1216  & 0.0440  & 0.1727  & 0.0824  & 0.1287  \\
Combined & DGS3   & 0.0256 & 0.1111  & 0.0374  & 0.1883  & 0.0855  & 0.1509  \\
Combined & DGS5   & 0.0189 & 0.0959  & 0.0346  & 0.2025  & 0.0770  & 0.1679  \\
Combined & DGS7   & 0.0145 & 0.0843  & 0.0329  & 0.2086  & 0.0698  & 0.1757  \\
Combined & DGS10  & 0.0126 & 0.0759  & 0.0327  & 0.2045  & 0.0633  & 0.1718  \\
Combined & DGS20  & 0.0121 & 0.0628  & 0.0367  & 0.1993  & 0.0507  & 0.1626  \\
Combined & DGS30  & 0.0129 & 0.0573  & 0.0414  & 0.1923  & 0.0444  & 0.1509  \\
\midrule
Combined & AAPL   & 0.0519 & 0.0621  & 5.6524  & 7.1124  & 0.0102  & 1.4600  \\
Combined & AMZN   & 0.0470 & 0.0617  & 5.1538  & 7.0294  & 0.0147  & 1.8756  \\
Combined & GOOGL  & 0.0399 & 0.0545  & 4.5781  & 6.0685  & 0.0146  & 1.4904  \\
Combined & MSFT   & 0.0318 & 0.0453  & 6.7621  & 10.1334 & 0.0135  & 3.3713  \\
Combined & NFLX   & 0.0786 & 0.0902  & 3.0877  & 3.4715  & 0.0116  & 0.3838  \\
Combined & NVDA   & 0.0761 & 0.0980  & 2.8012  & 3.3253  & 0.0219  & 0.5241  \\
Combined & TSLA   & 0.1114 & 0.1258  & 16.8843 & 20.7258 & 0.0144  & 3.8415 \\
\bottomrule
\end{tabular}
}
\end{table*}

\subsection{Leakage}

The approach taken in this analysis is susceptible to leakage critcisms, i.e., that the forecasts are already in the pre-training data for the model. One reason for using the Chronos class of models is that they were not trained on stock market and interest rate datasets. The development of Chronos-2 \citep{ansari_chronos-2_2025} emphasizes that for generalist pretrained models, the quality and diversity of training data are often more influential than the specific model architecture. While existing large-scale datasets provide a foundation, they are frequently limited to univariate series. To address this and foster robust in-context learning, Chronos-2 integrates established corpora like GIFT-Eval with a heavy reliance on synthetic data. This is achieved through two distinct methods: the TSI approach \citep{bahrpeyma_methodology_2021}, which blends randomized trend, seasonality, and noise components, and the TCM approach \citep{runge_causal_2023}, which uses temporal causal models and autoregression to generate complex, graph-based data patterns. The original Chronos univariate forecasting model \citep{ansari_chronos_2024} uses a large number of datasets, including some macroeconomic data from FRED \citep{godahewa_monash_2021} and some financial data in the datasets used in the Makridakis competition \citep{makridakis_m4_2020}. However, the datasets do not explicitly include any of the series we use for forecasting in this paper. And most certainly, the training data was not post-2023, so the results from that period do not suffer from leakage. 

\begin{figure*}
\centering
\includegraphics[width=\textwidth]{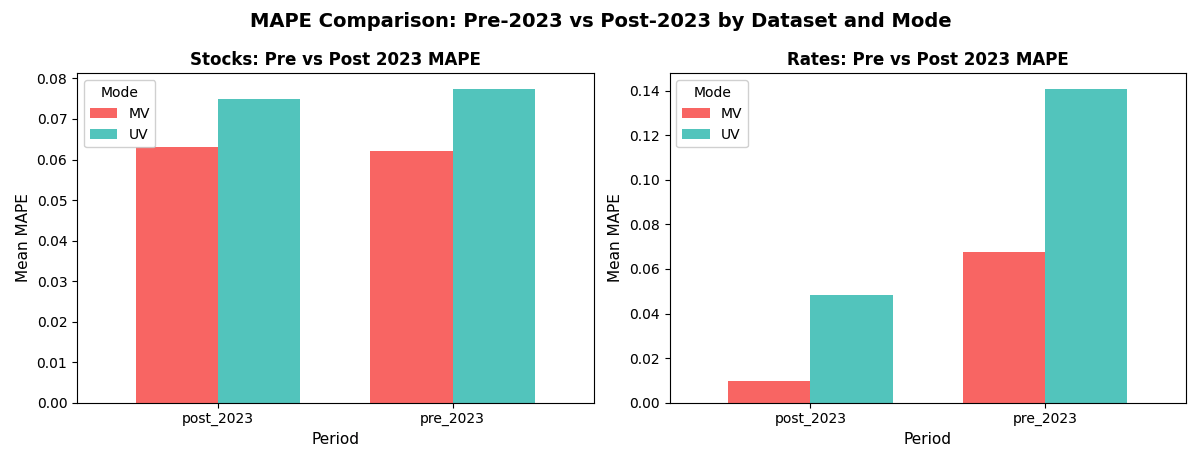}
\caption{\small Comparison of forecast accuracy pre-2023 and post-2023. Data used is for the period July 2010 through December 2025. }
\label{fig:pre_post_2023}
\end{figure*}

Nevertheless, if there were large amounts of leakage, because of training on time series that are in this paper or correlated to the ones used in this paper, then (i) we should see little difference between the MV and UV forecasts, and (ii) the pre-2023 forecasts should in fact be more accurate that the post-2023 ones, where the data is not likely to be correlated to or included in the pre-training dataset. First, we see that there is a great difference in the MV and UV forecasts. This is apparent in Tables \ref{tab:avg-performance}, \ref{tab:uv-mv-comparison}, and \ref{tab:uv-mv-combined}. With leakage there should be no material difference. Second, we look at the forecast accuracy pre-2023 and post-2023. Contrary to what we would expect with leakage, the post-2023 forecasts are more accurate, even though data from this period could not possibly be in the training data. This is shown for both rates and stocks in Figure \ref{fig:pre_post_2023}. This effect is more pronounced for rates than for stocks, where there is no clear difference. These results are unsupported if the model was contaminated by leakage from pre-training data. We conclude that the results are not impacted by leakage and that foundation models of this type do better when using multivariate series that come from the same contextual domain as the series that is being forecasted. 

\section{Concluding Discussion}

This paper evaluates Chronos-2 as a foundation-model approach to economic and financial forecasting, with a specific focus on whether multivariate inputs improve accuracy relative to univariate baselines. Across equities and U.S. Treasury rates, MV forecasts consistently outperform UV forecasts, with especially strong gains in the rates panel and meaningful improvements in equity forecasts, using MAPE and RMSE metrics. The results indicate that cross-series information from the same domain is valuable and that foundation models can capitalize on it when inputs are properly structured.

The findings also speak to the ongoing debate on the role of modern ML in financial forecasting. Consistent with prior evidence that classical benchmarks remain hard to beat, our results show that gains are not automatic; rather, they arise when models leverage multivariate structure and are evaluated with disciplined rolling protocols. This suggests that the most practical benefit of foundation models may come from their ability to transfer information across related series, rather than from raw model complexity alone. In addition,  foundation models can accommodate many input series, which is difficult to do with traditional time-series econometric models. 

There are several limitations. First, the study focuses on a single foundation model and a specific set of assets. Second, the evaluation uses fixed window and horizon choices; alternative regimes, higher-frequency data, or different rolling cadences may alter results. Third, while we report standard error metrics, practitioners may, depending on their specific use case, undertake further examination of tail risk, directional accuracy, and economic value to add depth to their analyses. 

Future work can extend the framework in three directions. One path is to broaden the universe to a larger ``world'' panel that includes macroeconomic, commodity, and international series to test transfer at scale. Another is to integrate uncertainty calibration, for example via conformal prediction, to improve the reliability of forecast intervals. Finally, hybrid designs that combine strong econometric baselines with foundation-model embeddings may offer robust performance while preserving interpretability.

\clearpage

\balance
\bibliographystyle{chicago}
\bibliography{ts}

\appendix

\section{Appendix: AI-Driven Workflow to Generate the Paper}
\label{ai_workflow}

This appendix outlines how the paper was produced with the assistance of Artificial Intelligence.

\section*{Detailed Workflow Step-by-Step}

Each step below was followed by human review, if AI was involved. 

\begin{enumerate}[label=\arabic*.]
    \item \textbf{Conceptualization:} Students are provided an initial outline.
    \item \textbf{Initial Coding:} Students prepare a Google Colab notebook with assistance from Gemini.
    \item \textbf{Refinement:} The professor updates the notebook, modularizing the code and updating logic, also utilizing Gemini for coding assistance.
    \item \textbf{Data Preparation:} Results are finalized in Colab, consisting of printed dataframes, plots, and textual outputs.
    \item \textbf{Environment Setup:} A \LaTeX{} shell for the paper is established in OpenAI Prism (\url{https://openai.com/prism/}).
    \item \textbf{Outlining:} The structure is set up with blank sections: Introduction, Methodology, Results, and Concluding Discussion (including relevant subsections).
    \item \textbf{Table Conversion:} Printed dataframes from Colab are fed into Gemini with prompts to convert them into \LaTeX{} tables. These are then integrated into the Results section.
    \item \textbf{Visuals:} Figures are added to the project folder, and the corresponding \LaTeX{} code for figure inclusion is written.
    \item \textbf{Drafting Results:} Prism is tasked with drafting text for the Results section to explain the tables and figures. Notably, the underlying model (GPT-5.2) inferred results directly from the tables to generate the write-up.
    \item \textbf{Methodology:} The Chronos-1 and Chronos-2 papers, along with experiment descriptions from the Colab notebook, are uploaded to Gemini. Gemini generates the Methodology section, which is then moved to the draft.
    \item \textbf{Specific Subsections:} A subsection on data leakage is added; the reasoning is provided by the authors while the LLM handles the prose.
    \item \textbf{Literature Review (Part 1):} A Bib\TeX{} file from Zotero is uploaded. Prism is prompted to generate a literature review based on the citations as they relate to the Methodology and Results.
    \item \textbf{Introduction and Literature Review (Part 2):} Previously published related papers are uploaded. Prism generates the Introduction and updates the literature review for context.
    \item \textbf{Conclusions:} Using the full paper as context, Prism generates the Concluding Discussion.
    \item \textbf{Abstract:} Prism is prompted to synthesize an abstract for the completed draft.
    \item \textbf{Human Editing:} Final human-led editing passes are performed to improve narrative flow, reformat the \LaTeX{} layout, and ensure technical accuracy.
\end{enumerate}

\subsection*{Self Review and Revision Process}
Gemini was also utilized to generate a mock referee report (\texttt{review.pdf}). This report informed a response document (\texttt{responses.md}) created with the following prompt: 
\begin{quote}
    \textit{``Please read the file ‘review.pdf‘ which is a referee report on the paper ‘chronos2.pdf‘ – then suggest changes and improvements to the paper in a document titled ‘response.md‘ that outlines the changes in response to cited sections of ‘review.pdf‘ as a response note to the review.''}
\end{quote}
Final edits were made to the manuscript based on these AI-suggested improvements. 

The code and files for the paper have been made public on GitHub: \url{https://github.com/srdas/timeseries-fm/tree/main}.

\end{document}